\documentstyle[12pt]{article}
\begin{document}
\def\beq{\begin{equation}}
\def\eeq{\end{equation}}
\def\bea{\begin{eqnarray}}
\def\eea{\end{eqnarray}}
\def\ve{\vert}
\def\vel{\left|}
\def\ver{\right|}
\def\nnb{\nonumber}
\def\ga{\left(}
\def\dr{\right)}
\def\aga{\left\{}
\def\adr{\right\}}
\def\rar{\rightarrow}
\def\nnb{\nonumber}
\def\la{\langle}
\def\ra{\rangle}
\def\lla{\left<}
\def\rra{\right>}
\def\ba{\begin{array}}
\def\ea{\end{array}}
\def\tep{$B \rar K \ell^+ \ell^-$}
\def\tepm{$B \rar K \mu^+ \mu^-$}
\def\tept{$B \rar K \tau^+ \tau^-$}
\def\ds{\displaystyle}
\def\es{\!\!\! &=& \!\!\!}
\def\ar{&+& \!\!\!}
\def\arr{\!\!\!\!&+&\!\!\!}
\def\ek{&-& \!\!\!}
\def\akk{\!\!\!\!&-&\!\!\!}
\def\cp{&\times& \!\!\!}
\def\se{\!\!\! &\simeq& \!\!\!}
\def\kpm{&\pm& \!\!\!}
\def\kmp{&\mp& \!\!\!}

% .........................................................

\def\simlt{\stackrel{<}{{}_\sim}}
\def\simgt{\stackrel{>}{{}_\sim}}

% .........................................................

\title{ {\Large {\bf 
Fourth generation effects in rare $B_s \rar \ell^+ \ell^-$ 
and $B_s \rar \ell^+ \ell^- \gamma$ decays 
        } 
                }       
      }

\author{\vspace{1cm}\\
{\small T. M. Aliev \thanks
{e-mail: taliev@metu.edu.tr}\,\,,
A. \"{O}zpineci \thanks
{e-mail: altugoz@metu.edu.tr}\,\,,
M. Savc{\i} \thanks
{e-mail: savci@metu.edu.tr}} \\
{\small Physics Department, Middle East Technical University} \\
{\small 06531 Ankara, Turkey} }
\date{}

\begin{titlepage}
\maketitle
\thispagestyle{empty}

\begin{abstract}
The contribution of the fourth generation quarks to the rare $B_s \rar \ell^+
\ell^-$ and $B_s \rar \ell^+ \ell^- \gamma$ decays are studied. 
Constraints on the extended Cabibbo--Kobayashi--Maskawa matrix elements
$\vel V_{tb} V_{ts}^\ast \ver$ and $\vel V_{t^\prime b} V_{t^\prime s}^\ast 
\ver$ are obtained from the existing experimental result for the branching 
ratio of the $B \rar X_s \gamma$ decay. 
The branching ratios of the
above--mentioned decays, as well as the photon spectrum in the 
$B_s \rar \ell^+ \ell^- \gamma$ decay, are studied. It is found that the
results are sensitive to the existence of the fourth generation.
\end{abstract}

%\vspace{1cm}
~~~PACS number(s): 12.60.--i, 13.20.--v, 13.20.He
\end{titlepage}

\section{Introduction}

The working and forthcoming high--statistics B--physics experiments at
BaBar, BELLE, HERA--B, the TeVatron and LHC--B \cite{R1} will probe the
flavor sector of the Standard Model (SM) with high precision. These
experiments may provide hints for the physics beyond the SM. Already, there have been
a large amount of theoretical investigations within various scenarios of new physics
to calculate the physical observables that will be measured.

The flavor--changing neutral current process $B_{s(d)} \rar \ell^+ \ell^-$
has attracted a lot of attention due to its sensitivity both to the gauge
structure of the SM and to the possible new physics beyond SM
\cite{R2}--\cite{R9}. Note that the experimental bound on the branching
ratio is \cite{R10}
\bea
&&{\cal B}(B_d \rar \mu^+ \mu^-) < 6.8 \times 10^{-7}~~~~~(CL=90\%)~,\nnb \\
&&{\cal B}(B_s \rar \mu^+ \mu^-) < 2.0 \times 10^{-6}~~~~~(CL=90\%)~.\nnb
\eea
In other words, measuring the branching ratio $B_{s(d)} \rar \ell^+ \ell^-$   
can impose the stringiest constraints on the possible parameter space of new
physics. The $B_s \rar \ell^+ \ell^- \gamma$ decay is also very sensitive to
the existence of the new physics beyond SM. In spite of the fact that this
decay has an extra $\alpha$ factor, it was shown in \cite{R11,R12} that the 
radiative leptonic $B_s \rar \ell^+ \ell^- \gamma~(\ell=e,~\mu)$ decays 
have larger decay rates 
compared to that purely leptonic ones, while for the $\tau$ lepton mode
widths for both decays are comparable to each other.

As has already been noted, the flavor--changing neutral current processes
$B_s \rar \ell^+ \ell^-$ and $B_s \rar \ell^+ \ell^- \gamma$ are very sensitive
to the new physics beyond the SM. One of the most straightforward and
economic extension of the SM is adding the fourth generation to the
fermionic sector, similar to the three generation case. 

Effects of the extra generation to the electroweak radiative correction and
its influence on the low energy physics processes were investigated in many
works (see for example \cite{R13} and references therein). 

In this work we study the effects of fourth generation in the 
$B_s \rar \ell^+ \ell^-$ and $B_s \rar \ell^+ \ell^- \gamma$ decays. The paper is
organized as follows. In section 2 we present theoretical description of the 
$B_s \rar \ell^+ \ell^-$ and $B_s \rar \ell^+ \ell^- \gamma$ decays for the
fourth generation case. Section 3 is devoted to the numerical analysis and
conclusion.
   
\section{Theoretical background}

In this section we will present the necessary theoretical expressions for
the $B_s \rar \ell^+ \ell^-$ and $B_s \rar \ell^+ \ell^- \gamma$ decays. 

Let us first consider $B_s \rar \ell^+ \ell^-$ decay. 
At quark level, this decay is described by the $b \rar s \ell^+ \ell^-$ 
transition for which the effective Hamiltonian can be written as
\bea
\label{effh}
{\cal H}_{eff} = \frac{\alpha G_F}{2 \sqrt{2} \pi} V_{tb} V_{ts}^\ast 
\sum_{i=1}^{10} C_i(\mu) {\cal O}_i(\mu)~, 
\eea
where the full set of operators in ${\cal O}_i(\mu)$ and the corresponding
expressions for the Wilson coefficients in the SM are given in
\cite{R14,R15}. As has been noted earlier, in the model we consider in the
present work, the fourth generation is introduced in the same way as the first
three generations are introduced in the SM. The fourth generation changes
only values of the Wilson coefficients $C_7(\mu),~C_9(\mu)$ and
$C_{10}(\mu)$ via the virtual exchange of the fourth generation up quark
$t^\prime$, i.e.,
\bea
\label{newc}
C_7^{tot}(\mu) &=& C_7^{SM}(\mu) + \frac{V_{t^\prime b}V_{t^\prime s}^\ast}
{V_{tb} V_{ts}^\ast} C_7^{new} (\mu) ~, \nnb \\
C_9^{tot}(\mu) &=& C_9^{SM}(\mu) + \frac{V_{t^\prime b}V_{t^\prime s}^\ast}
{V_{tb}V_{ts}^\ast} C_9^{new} (\mu) ~, \nnb \\
C_{10}^{tot}(\mu) &=& C_{10}^{SM}(\mu) + \frac{V_{t^\prime b}
V_{t^\prime s}^\ast}
{V_{tb}V_{ts}^\ast} C_{10}^{new} (\mu) ~,
\eea
where the last terms in these expressions correspond to the $t^\prime$ quark
contributions to the Wilson coefficients and $V_{t^\prime b}$ and
$V_{t^\prime s}$ are the elements of the $4\times 4$
Cabibbo--Kobayashi--Maskawa (CKM) matrix. The explicit forms of the
$C_i^{new}$ can easily be obtained from the
corresponding Wilson coefficient expressions in SM by simply substituting
$m_t \rar m_{t^\prime}$. The effective Hamiltonian (\ref{effh}) leads to the
following matrix element for the $b \rar s \ell^+ \ell^-$ transition
\bea
\label{mel}
{\cal M} &=& \frac{G\alpha}{2\sqrt{2} \pi}
 V_{tb}V_{ts}^\ast
\Bigg[ C_9^{tot} \, \bar s \gamma_\mu (1-\gamma_5) b \,
\bar \ell \gamma_\mu \ell +
C_{10}^{tot} \bar s \gamma_\mu (1-\gamma_5) b \,
\bar \ell \gamma_\mu \gamma_5 \ell \nnb \\
&-& 2  C_7^{tot}\frac{m_b}{q^2} \bar s \sigma_{\mu\nu} q^\nu
(1+\gamma_5) b \, \bar \ell \gamma_\mu \ell \Bigg]~,
\eea
where $q^2=(p_1+p_2)^2$ and $p_1$ and $p_2$ are the final leptons
four--momenta. It follows from Eq. (\ref{mel}) that in order to calculate
the matrix element for the $B_s \rar \ell^+ \ell^-$ decay, ${\cal M}$ is
sandwiched between vacuum and $B_s$ meson states, from which the matrix
elements $\lla 0 \vel \bar s \gamma_\mu (1 - \gamma_5) b \ver B_s \rra$ and
$\lla 0 \vel \bar s \sigma_{\mu\nu} q^\nu (1 + \gamma_5) b \ver B_s \rra$ need 
to be calculated. These matrix elements are equal to
\bea
\label{mtr}
\lla 0 \vel \bar s \gamma^\mu (1 - \gamma_5) b \ver B_s \rra &=& 
-i f_{B_s} p_B^\mu~, \nnb \\
\lla 0 \vel \bar s \sigma_{\mu\nu} q^\nu (1 + \gamma_5) b \ver B_s \rra &=& 0 ~.
\eea
Using these results, we get for the matrix element for 
$B_s \rar \ell^+ \ell^-$ decay 
\bea
\label{mel1}
{\cal M} = i \frac{G \alpha}{\sqrt{2} \pi} V_{tb} V_{ts}^\ast
m_\ell f_{B_s} C_{10}^{tot}(m_b) \, \bar \ell \gamma_5 \ell~.
\eea
Obviously, we see that the vector leptonic operator $\bar \ell \gamma_\mu
\ell$ does not contribute to $B_s \rar  \ell^+ \ell^-$ decay because it gives
zero when contracted with $B_s$ meson momentum.

Let us now turn our attention to the $B_s \rar \ell^+ \ell^- \gamma$ decay.
The matrix element for the $b \rar s \ell^+ \ell^- \gamma$ decay can be
obtained from that of the $b \rar s \ell^+ \ell^-$. For this purpose it is
necessary to attach the photon to any charged internal as well as external
line. When the photon is attached to internal lines, there will be a
suppression factor of $m_b^2/m_W^2$ in the Wilson coefficient, since the
resulting operators (dimension--8) are two order higher dimensionally than
the usual ones (dimension--6). Thus the main contributions to the
$b \rar s \ell^+ \ell^- \gamma$ decay come from diagrams when photon is 
radiated from initial (structure dependent SD part) and final
(internal Bremsstrahlung IB part) fermions.

In order to calculate the matrix element corresponding to SD part, the
following matrix elements must be calculated in the first hand 
\bea
\label{ndd}
&&\lla \gamma \vel \bar s \gamma_\mu (1-\gamma_5) b \ver B_s \rra~, \nnb \\
&&\lla \gamma \vel \bar s \sigma_{\mu\nu} q^\nu (1+\gamma_5) b \ver B_s \rra~. \nnb
\eea
These matrix elements can be parametrized in
terms of two independent, gauge invariant, parity conserving and parity
violating form factors \cite{R11,R16} as 
\bea
\label{mel2}
\lla \gamma \vel \bar q \gamma_\mu (1-\gamma_5) b \ver B_s \rra &=&
\frac{e}{m_{B_s}^2} \Big\{ \epsilon_{\mu\nu\lambda\sigma} \varepsilon^{\ast\nu} 
q^\lambda k^\sigma g(q^2) + i \Big[ \varepsilon_\mu^\ast (kq) -
(\varepsilon^\ast q) k_\mu \Big] f(q^2) \Big\}~,\\
\label{mel3}
\lla \gamma \vel \bar s i \sigma_{\mu\nu} q^\nu (1+\gamma_5) b \ver B_s 
\rra &=& \frac{e}{m_{B_s}^2}\Big\{ \epsilon_{\mu\nu\lambda\sigma}
\varepsilon^{\ast\nu}    
q^\lambda k^\sigma g_1(q^2) + i \Big[ \varepsilon_\mu^\ast (kq) -
(\varepsilon^\ast q) k_\mu \Big] f_1(q^2) \Big\}~,
\eea
where $\varepsilon^\mu$ and $k^\mu$ are the four--vector polarization and
four--momentum of the photon, respectively, and $q^\mu$ is the momentum
transfer. Using Eqs. (\ref{mel}), (\ref{mel2}) and (\ref{mel3}), matrix element 
describing the SD part takes the following form
\bea
\label{msd}
{\cal M}_{SD} &=& \frac{\alpha G_F}{2 \sqrt{2} \pi} \frac{e}{m_{B_s}^2} 
V_{tb} V_{ts}^\ast \Big\{\epsilon_{\mu\nu\lambda\sigma}
\varepsilon^{\ast\nu} q^\lambda k^\sigma \Big[A_1 \bar \ell \gamma^\mu \ell
+ A_2 \bar \ell \gamma^\mu \gamma_5 \ell \Big]\nnb \\
&+& i \Big[ \varepsilon_\mu^\ast (kq) -(\varepsilon^\ast q) k_\mu \Big]
\Big[B_1 \bar \ell \gamma^\mu \ell+ B_2 \bar \ell \gamma^\mu \gamma_5 \ell
\Big]\Big\}~,
\eea    
where
\bea
A_1 &=& C_9^{tot} g(q^2) - 2 C_7^{tot} \frac{m_b}{q^2} g_1(q^2) ~, \nnb \\
A_2 &=& C_{10}^{tot} g(q^2)~, \nnb \\
B_1 &=& C_9^{tot} f(q^2) - 2 C_7^{tot} \frac{m_b}{q^2} f_1(q^2)~, \nnb \\
B_2 &=& C_{10}^{tot} f(q^2)~. \nnb
\eea
When a photon is radiated from the final leptons, the corresponding matrix
element is
\bea
\label{mib}
{\cal M}_{IB} = \frac{\alpha G_F}{\sqrt{2} \pi} V_{tb} V_{ts}^\ast e f_{B_s}
m_\ell  C_{10}^{tot} \Bigg[ \bar \ell \Bigg( 
\frac{{\not\!\varepsilon}^\ast{\not\!p}_{B_s}}{2 p_1 k} -
\frac{{\not\!p}_{B_s}{\not\!\varepsilon}^\ast}{2 p_2 k} \Bigg) 
\gamma_5 \ell \Bigg]~.
\eea
Finally the total matrix element for the $B_s \rar \ell^+ \ell^- \gamma$ decay
is equal to a sum of the Eqs. (\ref{msd}) and (\ref{mib})
\bea
{\cal M} = {\cal M}_{SD} + {\cal M}_{IB}~.\nnb
\eea
The double differential decay width of the $B_s \rar \ell^+ \ell^- \gamma$
process, in the rest frame of the $B_s$ meson, is
\bea
\label{dddw}
\frac{d\Gamma}{dE_\gamma \,dE_1} = \frac{1}{256 \pi^3 m_{B_s}} \vel \overline {\cal M}
\ver^2~,
\eea
where $E_\gamma$ and $E_1$ are the energies of the photon and one of the final
leptons, respectively, bar means that summation over spin of final
particles is performed. 
The physical regions of $E_\gamma$ and $E_1$ are determined from the following
inequalities
\bea
0 \leq E_\gamma \leq \frac{m_{B_s}^2 - 4 m_\ell^2}{2 m_{B_s}}~,\nnb
\eea
\bea
\frac{m_{B_s} - E_\gamma}{2} - \frac{E_\gamma}{2} v \leq E_1 \leq
\frac{m_{B_s} - E_\gamma}{2} + \frac{E_\gamma}{2} v ~,
\eea
where 
\bea
v=\sqrt{1 - \frac{4 m_\ell^2}{q^2}}~,\nnb
\eea
is the lepton velocity.
The $\vel {\cal M}_{SD} \ver^2$ term is infrared divergence free, the 
interference term $2 \mbox{\rm Re}\ga {\cal M}_{SD}{\cal M}_{IB}^\ast\dr$ has
integrable infrared singularity and only $\vel {\cal M}_{IB} \ver^2$ has
infrared singularity due to the emission of soft photon. In the soft photon
limit the $B_s \rar  \ell^+ \ell^- \gamma$ decay cannot be distinguished from
the pure leptonic $B_s \rar  \ell^+ \ell^-$ decay. Therefore,
in order to obtain a finite result the $B_s \rar \ell^+ \ell^- \gamma$ and the 
pure leptonic $B_s \rar \ell^+ \ell^-$ decay with radiative
corrections must be considered together. It was shown in \cite{R17} that 
when both processes are considered together,
all infrared singularities coming from the real photon emission and the
virtual photon corrections are indeed canceled.
However in the present work our point of view is slightly different,
namely, instead of following the above mentioned procedure, we consider the 
$B_s \rar \ell^+ \ell^- \gamma$ decay as a separate process but not as the 
${\cal O}(\alpha)$
correction to the $B_s \rar \ell^+ \ell^-$ decay. In other words, we consider
the photon in the $B_s \rar \ell^+ \ell^- \gamma$ decay as a hard and
detectable one. Therefore, in order to obtain the decay width we impose a 
cut on the photon energy, which will correspond to the experimental cut 
imposed on the minimum energy for the detectable photon. We require the photon 
energy to be larger than $25~MeV$, i.e., $E_\gamma \ge \delta m_{B_s}/2$, 
where $\delta \ge 0.010$.

Using Eqs. (\ref{mel1}) and (\ref{dddw}) we get the following results for the 
$B_s \rar \ell^+ \ell^-$ and $B_s \rar \ell^+ \ell^- \gamma$ decays, respectively:
\bea
\label{bll}
\lefteqn{
{\cal B} (B_s \rar \ell^+ \ell^-) = \frac{G_F^2 \alpha^2}{64 \pi^2} m_{B_s}^3
\tau_{B_s} f_{B_s}^2 \vel V_{tb} V_{ts}^\ast \ver^2 
\frac{4 m_\ell^2}{m_{B_s}^2} \sqrt{1-\frac{4
m_\ell^2}{m_{B_s}^2}} \vel C_{10}^{tot}\ver^2}~. \\
\label{bllg}
\lefteqn{   
{\cal B} (B_s \rar \ell^+ \ell^- \gamma) = \vel \frac{G_F
\alpha}{2\sqrt{2}\pi} \ver^2 \vel V_{tb} V_{ts}^\ast \ver^2
\frac{\alpha \pi}{(2 \pi)^3} m_{B_s}^3 \tau_{B_s}} \nnb \\
&&\times\,\Bigg\{ \int_0^{1-4 r} dx \, x^3 v\Bigg[ \ga \vel A_1 \ver^2 + 
\vel B_1 \ver^2 \dr (1-x+2 r) + \ga \vel A_2 \ver^2 + 
\vel B_2\ver^2 \dr (1-x-4 r) \Bigg] \nnb \\
&&+\, 2 r f_{B_s} C_{10}^{tot} \int_0^{1-4 r} dx \,
x^2 \, \mbox{\rm Re} \ga A_1 \dr {\rm ln} \frac{1 + v}{1 - v}\nnb \\
&&-\,4 \vel f_{B_s} C_{10}^{tot}\ver^2 r  \int_\delta^{1-4 r} dx \,
\Bigg[ 2 v \frac{1-x}{x} + \ga 2 + \frac{4 r}{x} - \frac{2}{x} -x \dr 
{\rm ln} \frac{1 + v}{1 - v} \Bigg] \Bigg\}~,
\eea   
where $x=2 E_\gamma/m_{B_s}$ is the dimensionless photon energy and
$r=m_\ell^2/m_{B_s}^2$.

It follows from Eq. (\ref{bllg}) that to be able to calculate the decay
width, the
explicit forms of the form factors $g,~f,~g_1$ and $f_1$ are needed. 
In further analysis, use will be made of the form factors that are predicted
by the light--cone $QCD$ sum rules
\cite{R11,R16}, whose $q^2$ dependences, to a very good
accuracy,
can be written in the following dipole forms,
\bea
\label{ff}
g_(q^2) &=& \frac{1~GeV}{\ds \ga 1-\frac{q^2}{(5.6~GeV)^2}\dr^2}~,
~~~~~~~~~~
f_(q^2) = \frac{0.8~GeV}{\ds \ga 1-\frac{q^2}{(6.5~GeV)^2}\dr^2}~,
\nnb \\ \nnb \\ 
g_1(q^2) &=& \frac{3.74~GeV^2}{\ds \ga 1-\frac{q^2}{40.5~GeV^2}\dr^2}~,
~~~~~~~~~    
f_1(q^2) = \frac{0.68~GeV^2}{\ds \ga 1-\frac{q^2}{30~GeV^2}\dr^2}~.
\eea
As is obvious from Eq. (\ref{newc}), to get numerical values of the
branching ratios the value of the fourth generation CKM matrix element
$\vel V_{t^\prime b} V_{t^\prime s}^\ast\ver$ is needed. For this purpose we will
use the experimentally measured values of the branching ratios 
${\cal B}(B \rar X_s \gamma)$ and ${\cal B}(B \rar X_c e \bar \nu_e)$. 
In order to reduce the uncertainties coming from $b$ quark
mass, we consider the ratio
\bea
R = \frac{{\cal B}(B \rar X_s \gamma)}{{\cal B}(B \rar X_c e \bar \nu_e)}~.
\label{altug}
\eea
In leading logarithmic approximation this ratio can be written as
\bea
\label{rrr}
R = \frac{6 \alpha \vel C_7^{tot}(m_b)V_{tb} V_{ts}^\ast \ver^2}
{\pi f(\hat m_c) \kappa(\hat m_c) \vel V_{cb} \ver^2}
~,
\eea
where $\hat m_c = m_c/m_b$ and the phase factor $f(\hat m_c)$ and ${\cal O}(\alpha_s)$ 
QCD correction
factor $\kappa(\hat m_c)$ \cite{R18} of $b \rar c \ell \bar \nu$
are given by
\bea
\label{fff}
f(\hat m_c)&=& 1 - 8 \hat m_c^2 + 8 \hat m_c^6 - \hat m_c^8 -
24 \hat m_c^4 \,\mbox{\rm ln} (\hat m_c) ~, \nnb \\ \nnb \\
\label{kap}
\kappa(\hat m_c)&=& 1 - \frac{2 \alpha_s(m_b)}{3 \pi}
\Bigg[ \ga \pi^2 - \frac{31}{4} \dr \ga 1 - \hat m_c^2 \dr^2 + \frac{3}{2}
\Bigg]~.
\eea
From Eqs. (\ref{rrr}) and (\ref{altug})  we get
\bea
\label{vtb}
\vel C_7^{SM} V_{tb} V_{ts}^\ast + 
C_7^{new} V_{t^\prime b} V_{t^\prime s}^\ast \ver =
\sqrt{ 
\frac{\pi f(\hat m_c) \kappa(\hat m_c) \vel V_{cb} \ver^2}{6 \alpha}~
\frac{{\cal B}(B \rar X_s \gamma)}{{\cal B}(B \rar X_c e \bar \nu_e)}}~.
\eea   

The model parameters can be constrained from the measured branching ratio of
the $B \rar X_s \gamma$ decay
\bea
{\cal B}(B \rar X_s \gamma) = \left\{ \begin{array}{lc}
~~~\left( 3.21 \pm 0.43 \pm 0.27^{+0.18}_{-0.10} \right) \times
10^{-4}& \cite{R19}\\ \\
~~~\left( 3.36 \pm 0.53 \pm 0.42 \pm 0.54 \right) \times
10^{-4}& \cite{R20}\\ \\ 
~~~\left( 3.11 \pm 0.80 \pm 0.72\right) \times
10^{-4}& \cite{R21}\end{array} \right. \nnb
\eea

In further numerical analysis, we will use the weighted average value 
${\cal B}(B \rar X_s \gamma) = (3.23 \pm 0.42)\times 10^{-4}$ \cite{R22}
for the branching ratio of the $B \rar X_s \gamma$ decay.

Next constraint to the extended CKM matrix element comes from the
unitarity condition, i.e., 
\bea
\label{e18}
\vel V_{us} \ver^2 + \vel V_{cs} \ver^2 + \vel V_{ts} \ver^2 +
\vel V_{t^\prime s} \ver^2 = 1~,\nnb \\
\vel V_{ub} \ver^2 + \vel V_{cb} \ver^2 + \vel V_{tb} \ver^2 +
\vel V_{t^\prime b} \ver^2 = 1~,\nnb \\
V_{ub} V_{us}^\ast + V_{cb} V_{cs}^\ast + V_{tb} V_{ts}^\ast +
V_{t^\prime b} V_{t^\prime s}^\ast = 0 ~.
\eea

At this point we would like to make the following remark: Charged--current
three--level decays are well measured experimentally and they are not
affected by new physics at leading order.  For this reason, in the following
discussions  we will use
Particle Data Group (PDG) constraints \cite{R10} for $\vel V_{ud} \ver$, 
$\vel V_{us} \ver$, $\vel V_{cs} \ver$, $\vel V_{cb} \ver$ and 
$\vel V_{ub}/V_{cb} \ver$. Using the weighted average for ${\cal B} (B \rar
X_s \gamma)$ and PDG constraint $0.38 \le \vel V_{cb} \ver \le 0.044$
\cite{R10}, from Eqs. (\ref{vtb}) and (\ref{e18}) we obtain the following 
constraints
\bea 
\label{e19}
0.011 \le \vel C_7^{SM} V_{tb} V_{ts}^\ast \right. \arr \left. C_7^{new} V_{t^\prime b} 
V_{t^\prime s}^\ast \ver \le 0.015~,\\
\label{e20}
0.03753 \le \vel V_{tb} V_{ts}^\ast \right. \arr \left. V_{t^\prime b} V_{t^\prime s}^\ast
\ver \le 0.043976~,\\
\label{e21}
0 \le \vel V_{ts} \ver^2 \arr \vel V_{t^\prime s} \ver^2 \le 0.00492~,\\
\label{e22}
0.998 \le \vel V_{tb} \ver^2 \arr  \vel V_{t^\prime b} \ver^2 \le 0.9985~.
\eea  

\section{Numerical analysis}

In this section we study the constraint on $V_{t^\prime b} V_{t^\prime
s}^\ast$ and using the resulting bound on $\vel V_{t^\prime b} V_{t^\prime
s}^\ast \ver$ we investigate the influence of fourth generation on the
branching ratios. Before performing this analysis we present the values of
the main input parameters which we use in our calculations:
$m_b = 4.8~GeV,~m_c=1.35~GeV,~m_{B_s}=5.369~GeV,~\tau_{B_s} = 1.64\times 10^{-12}~s$. 
Furthermore in calculating the
branching ratios the values of the Wilson coefficients $C_7^{SM}(m_b),~
C_9^{SM}(m_b)$ and $C_{10}^{SM}(m_b)$ are needed. 
In leading logarithmic approximation
$C_7^{SM}(m_b)=-0.315$ and $C_{10}^{SM}(m_b)=-4.642$ (see \cite{R14,R15}).
The analytic expression of $C_9^{SM}(m_b)$ is given as
\bea
\label{c9}
C_9^{SM}(m_b) &=& 4.227 + g(\hat m_c, \hat s)
(3 C_1 + C_2 + 3 C_3 + C_4 + 3 C_5 + C_6) \nnb \\
&-& \frac{1}{2} g(0, \hat s)
(C_3 + 3 C_4) - \frac{1}{2} g(1, \hat s) (4 C_3 + 4 C_4 + 3 C_5 +
C_6)\nnb \\
&+& \frac{2}{9} (3 C_3 + C_4 + 3 C_5 + C_6)~,\nnb
\eea
where $\hat s = q^2/m_b^2$. The explicit form of the expressions for the
functions $g(\hat m_i, \hat s)$ and the values of the individual Wilson coefficients
can be found in \cite{R14,R15}.

It should be noted that $B_s \rar \ell^+ \ell^- \gamma$ can also receive long
distance contributions which have their origin in the real intermediate
states, i.e., $J/\psi,~\psi^\prime,~\cdots$. In the present work we neglect
such long distance contributions. 

In order to study the influence of the fourth generation quarks to the rare
$B_s \rar \ell^+ \ell^-$ and $B_s \rar \ell^+ \ell^- \gamma$ decays we need the
values of $V_{tb} V_{ts}^\ast$ and $V_{t^\prime b} V_{t^\prime s}^\ast$. For
this purpose we carry out the following analysis. We consider $V_{tb}
V_{ts}^\ast$ and $V_{t^\prime b} V_{t^\prime s}^\ast$ to be two independent
complex parameters. They are constrained by the unitarity condition ({\ref{e18})
and the measured experimental value on branching ratio of the $B \rar X_s
\gamma$ decay (see Eq. (\ref{e19})) which depends on $m_{t^\prime}$. 
Thus for each value of
$m_{t^\prime}$ there exists an allowed region in the 
($\vel V_{tb} V_{ts}^\ast\ver$ and $\vel V_{t^\prime b} V_{t^\prime
s}^\ast\ver$) plane. 

It is not possible to solve the constraints Eqs. (\ref{e19})-(\ref{e22}) analytically.
Therefore, we chose a large number of random {\em complex} values for 
$V_{tb}$, $V_{ts}$, $V_{t'b}$, $V_{t's}$ such that the constraints are all satisfied. 
For a sufficiently large number, the selected values would range over the whole solution space.
In Figs. (1)--(3), to give an idea about the solution space, we present
the allowed region in the plane of the absolute values of $V_{tb} V_{ts}^\ast$ and 
$V_{t^\prime b} V_{t^\prime s}^\ast$ at $m_{t^\prime}=200~GeV$, $m_{t^\prime}=300~GeV$, and $m_{t^\prime}=300~GeV$, respectively.
The upper left boundary of the allowed region is determined by Eq. (\ref{e21}) and (\ref{e22}) whereas the
other three boundaries are determined by Eqs. (\ref{e19}) and (\ref{e20}).
From these figures it can easily
be seen that, when $\vel V_{t^\prime b} V_{t^\prime s}^\ast\ver \simeq 0$
the value of $\vel V_{tb} V_{ts}^\ast\ver $ is close to its SM value and is
evenly distributed around 0.04. However, when the value of $\vel V_{t^\prime b} V_{t^\prime s}^\ast\ver$ 
gets larger, up to a critical value around $\vel V_{t^\prime b} V_{t^\prime s}^\ast\ver \sim 0.012$, 
the allowed region for $\vel V_{tb} V_{ts}^\ast\ver$ becomes wider with the center fixed at 0.04. Beyond the critical point, 
the center of the region shifts toward smaller values with its width fixed. This behavior of the allowed range for
$\vel V_{tb} V_{t^s}^\ast\ver$ continues until $\vel V_{t^\prime b} V_{t^\prime s}^\ast\ver$ reaches
a second critical value around $\vel V_{t^\prime b} V_{t^\prime s}^\ast\ver \sim
0.04$.
When $\vel V_{t^\prime b} V_{t^\prime s}^\ast\ver$ takes values beyond this value,  the
allowed region for $\vel V_{tb} V_{ts}^\ast\ver$ begins to shrink to zero with its center fixed around 0.012.
Similar behavior is observed for all other values of $m_{t^\prime}$. After
determining the physical regions for $V_{tb} V_{ts}^\ast$
and $V_{t^\prime b} V_{t^\prime s}^\ast$, the range of the region for the branching ratio of 
the $B_s \rar \ell^+ \ell^-$  and $B_s \rar \ell^+ \ell^- \gamma$ 
decays can be determined.       
Depicted in Figs. (4) and (5) are the dependence of the ranges of the branching ratios of
the $B_s \rar \mu^+ \mu^-$ and $B_s \rar \tau^+ \tau^-$ decays on
$m_{t^\prime}$, respectively. It is evident from both figures that at lower
values of $m_{t^\prime}$ ($200~GeV \le m_{t^\prime} \le 250~GeV$) the
branching value is distributed mainly around the SM value. 
This is to be expected since when $m_{t'}=m_t$, the contributions to the Wilson
coefficients from the new physics effects are identical to those from the SM. In particular, after dividing
Eq. (\ref{e19}) by $\vel C_7^{SM} \ver$, Eq. (\ref{e19}) becomes 
a constraint on $\vel V_{tb} V_{ts}^\ast +  V_{t^\prime b} V_{t^\prime s}^\ast \ver$ just as Eq. (\ref{e20}), and 
the branching ratio itself also depends on this combination only.
Obviously, for larger values of $m_{t^\prime}$ the possible departure from the SM
prediction becomes substantial, signaling the existence of new physics
beyond SM.

It follows from these figures that there  exists a region for $V_{tb}
V_{ts}^\ast$ and $V_{t^\prime b} V_{t^\prime s}^\ast$ where branching ratio
attains smaller values than the SM prediction for all values of
$m_{t^\prime}$. This can be explained as follows. It follows from Eq.
(\ref{bll}) that the branching ratio is proportional to 
\bea
{\cal B}(B_s \rar \ell^+ \ell^-) \propto \vel C_{10}^{tot} \ver^2 \vel V_{tb}
V_{ts}^\ast \ver^2 \equiv
\vel C_{10}^{SM} \ver^2 \vel 1+ \frac{V_{t^\prime b} V_{t^\prime s}^\ast}
{V_{tb} V_{ts}^\ast} \, \frac{C_{10}^{new}}{C_{10}^{SM}} \ver^2 
\vel V_{tb} V_{ts}^\ast \ver^2~. \nnb 
\label{altug2}
\eea
The ratio $C_{10}^{new}/C_{10}^{SM}$ is positive for all values of
$m_{t^\prime}$ and increases with increasing $m_{t^\prime}$. 
From Eq. (\ref{altug2}), we see that depending on whether 
$\vel 1 + \lambda  \ver$, where 
\bea
\lambda = \frac{V_{t^\prime b} V_{t^\prime s}^\ast}
{V_{tb} V_{ts}^\ast} \, \frac{C_{10}^{new}}{C_{10}^{SM}}~,\nnb 
\eea 
is greater then, 
equal to, or less then one, the branching ration is greater then, equal to, or less then 
the SM value. The boundary separating these regions, $|\lambda + 1| = 1$ can be considered as
a circle on the complex $\lambda$ plane with radius $1$ and center at $-1$. Within this circle,
the predicted branching ratio is smaller then the SM value, whereas out of this circle, it is 
bigger.
%Put in other words, if
%\bea\frac{\vel V_{t^\prime b} V_{t^\prime s}^\ast\ver }
%{\vel V_{tb} V_{ts}^\ast \ver} > 2 \frac{C_{10}^{SM}}{C_{10}^{new}}~, \nnb
%\eea
%holds then the resulting branching ratio is larger compared to that of the
%SM prediction. On the other case, if
%\bea
%0 < \frac{\vel V_{t^\prime b} V_{t^\prime s}^\ast\ver }
%{\vel V_{tb} V_{ts}^\ast \ver} < 2 \frac{C_{10}^{SM}}{C_{10}^{new}}~, \nnb 
%\eea
%then the branching ratio can be bigger or smaller than the SM result depending 
%on the relative phase between $V_{t^\prime b} V_{t^\prime s}^\ast$ and 
%$V_{tb} V_{ts}^\ast$.

The dependence of the branching ratio on $m_{t^\prime}$ for the $B_s
\rar \mu^+ \mu^- \gamma$ and $B_s \rar \tau^+ \tau^- \gamma$ decays 
are shown in Figs. (6) and (7). Qualitative understanding of the dependence
of the branching ratio on $m_{t^\prime}$ for the $B_s \rar \ell^+ \ell^-
\gamma~(\ell=\mu,~\tau)$ is analogous to the 
$B_s \rar \ell^+ \ell^-~(\ell=\mu,~\tau)$ case. 
The lower bound of $x$ is chosen to be $x=0.01$ 
in order to get finite result for the ${\cal B}((B_s \rar \ell^+ \ell^- \gamma)$
decay (see Eq. (\ref{bllg})).
It should be noted here
that the branching ratio for the $B_s \rar \mu^+ \mu^- \gamma$ decay
practically is insensitive to the variation of $x_{min}$. For example when
$x_{min}$ is varied from $10^{-3}$ to $10^{-1}$, the branching ratio
changes only about $\sim 5\%$.  

Our final remark is that fourth generation can give contribution to the rare
$B$ and $K$ decays. For this reason constraints on fourth generation
imposed by the
$\rho$ parameter, $\Delta m_K,~\varepsilon_K,~B_{d(s)}^0 - \bar B_{d(s)}^0$
mixing, $K_L \rar \pi^0 \nu \bar \nu$, $K_L \rar \mu^+ \mu^-$ decays, must
be considered together with $b \rar s \gamma$ decay. Early attempts in this
direction were done in \cite{R22}. A more detailed analysis of constraints
arising from rare $B$ and $K$ meson decays is presently under study.    

In conclusion, we have studied the effect of the fourth generation quarks to
the rare $B_s \rar \ell^+ \ell^-$ and $B_s \rar \ell^+ \ell^- \gamma$ decays.
From the experimental result for the branching ratio of $B \rar X_s \gamma$
decay, the values of $\vel V_{tb} V_{ts}^\ast \ver$ and 
$\vel V_{t^\prime b} V_{t^\prime s}^\ast \ver$ are constrained. We have
observed that, depending on the relative phase between $V_{tb}
V_{ts}^\ast$ and $ V_{t^\prime b} V_{t^\prime s}^\ast$, fourth
generation quarks can enhance or suppress the 
SM result. As a result the branching ratios of $B_s \rar \ell^+ \ell^-$ and
$B_s \rar \ell^+ \ell^- \gamma$ decays can be larger or smaller compared to
that of the SM prediction.

\newpage

\section*{Figure captions}
{\bf Fig. (1)} The allowed regions for $\vel V_{tb} V_{ts}^\ast \ver$ and
$\vel V_{t^\prime b} V_{t^\prime s}^\ast \ver$ at $m_{t^\prime}=200~GeV$.\\ \\
{\bf Fig. (2)} The same as in Fig. (1), but at $m_{t^\prime}=300~GeV$.\\ \\ 
{\bf Fig. (3)} The same as in Fig. (1), but at $m_{t^\prime}=400~GeV$.\\ \\
{\bf Fig. (4)} The dependence of the branching ratio for the
$B_s \rar \mu^+ \mu^-$ decay on the fourth up quark mass 
$m_{t^\prime}$.\\ \\
{\bf Fig. (5)} The same as in Fig. (4), but for the $B_s \rar \tau^+ \tau^-$
decay.\\ \\
{\bf Fig. (6)} The same as in Fig. (4), but for the 
$B_s \rar \mu^+ \mu^-\gamma $ decay.\\ \\
{\bf Fig. (7)} The same as in Fig. (4), but for the $B_s \rar \tau^+ \tau^-
\gamma$ decay.\\ \\

\newpage

\begin{figure}
\vskip 1.5 cm
    \includegraphics{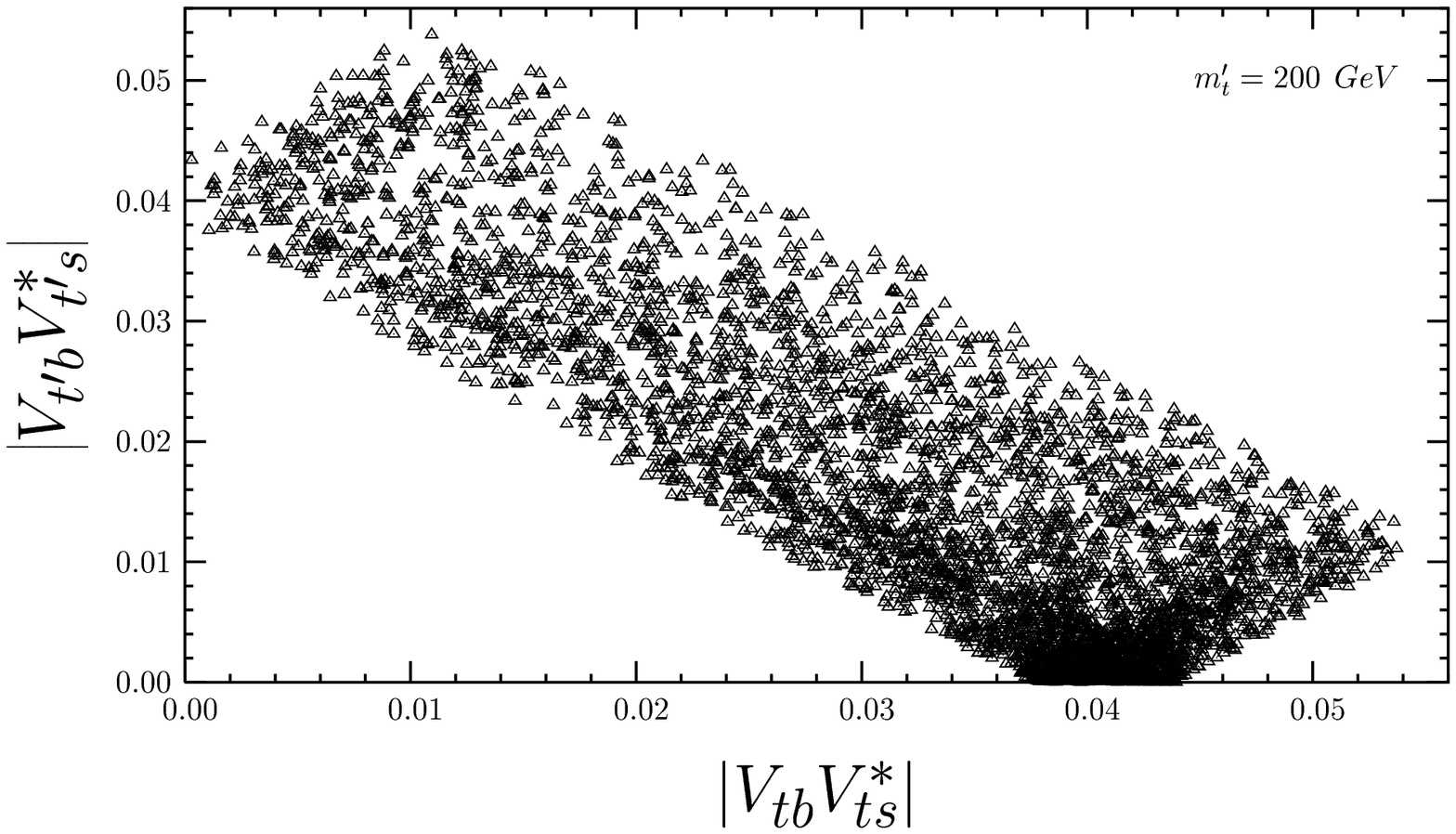}
\vskip 7.8cm
\caption{}
\end{figure}

\begin{figure}
\vskip 2.5 cm
    \includegraphics{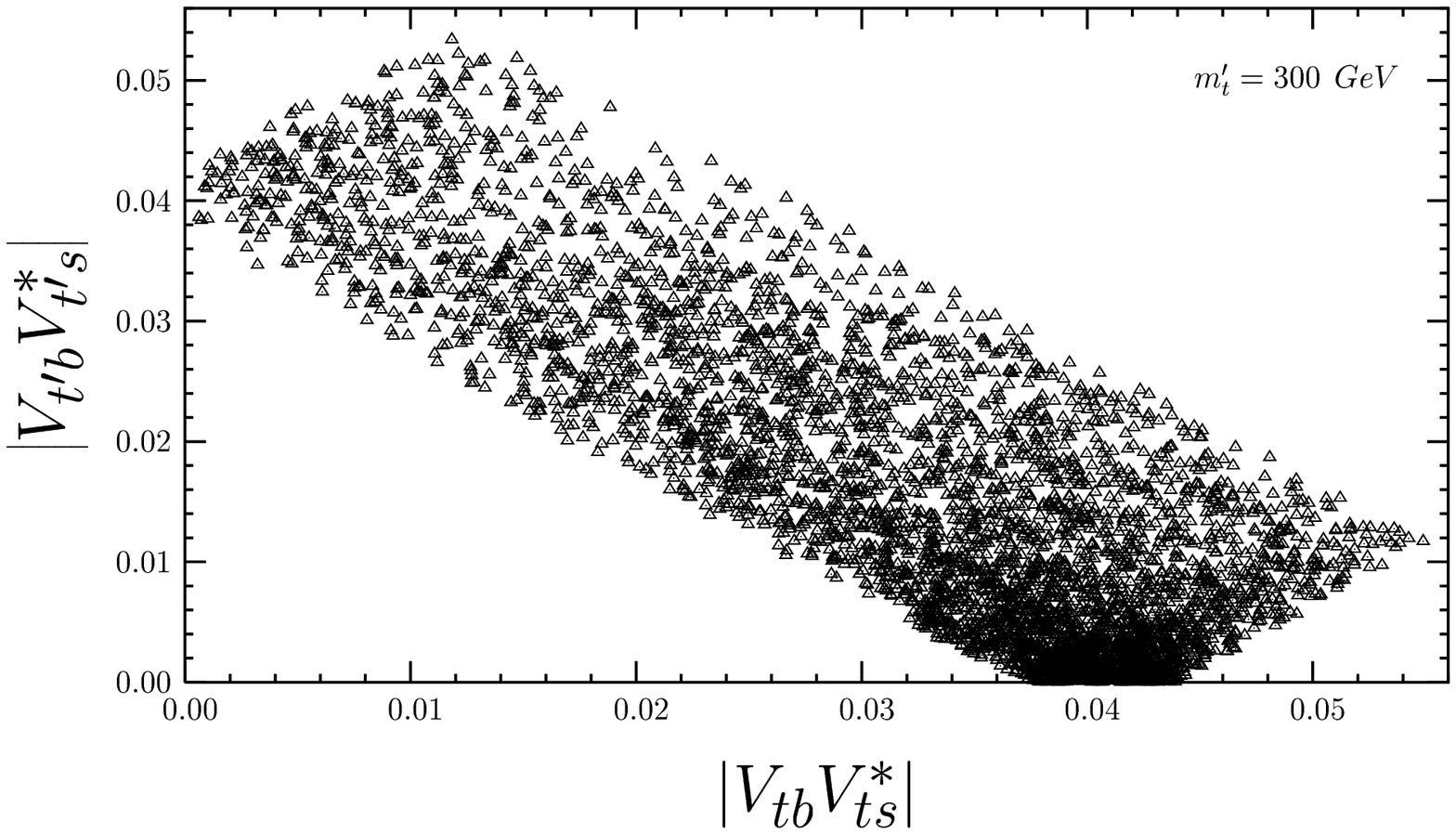}
\vskip 7.8 cm
\caption{}
\end{figure}

\begin{figure}
\vskip 1.5 cm
    \includegraphics{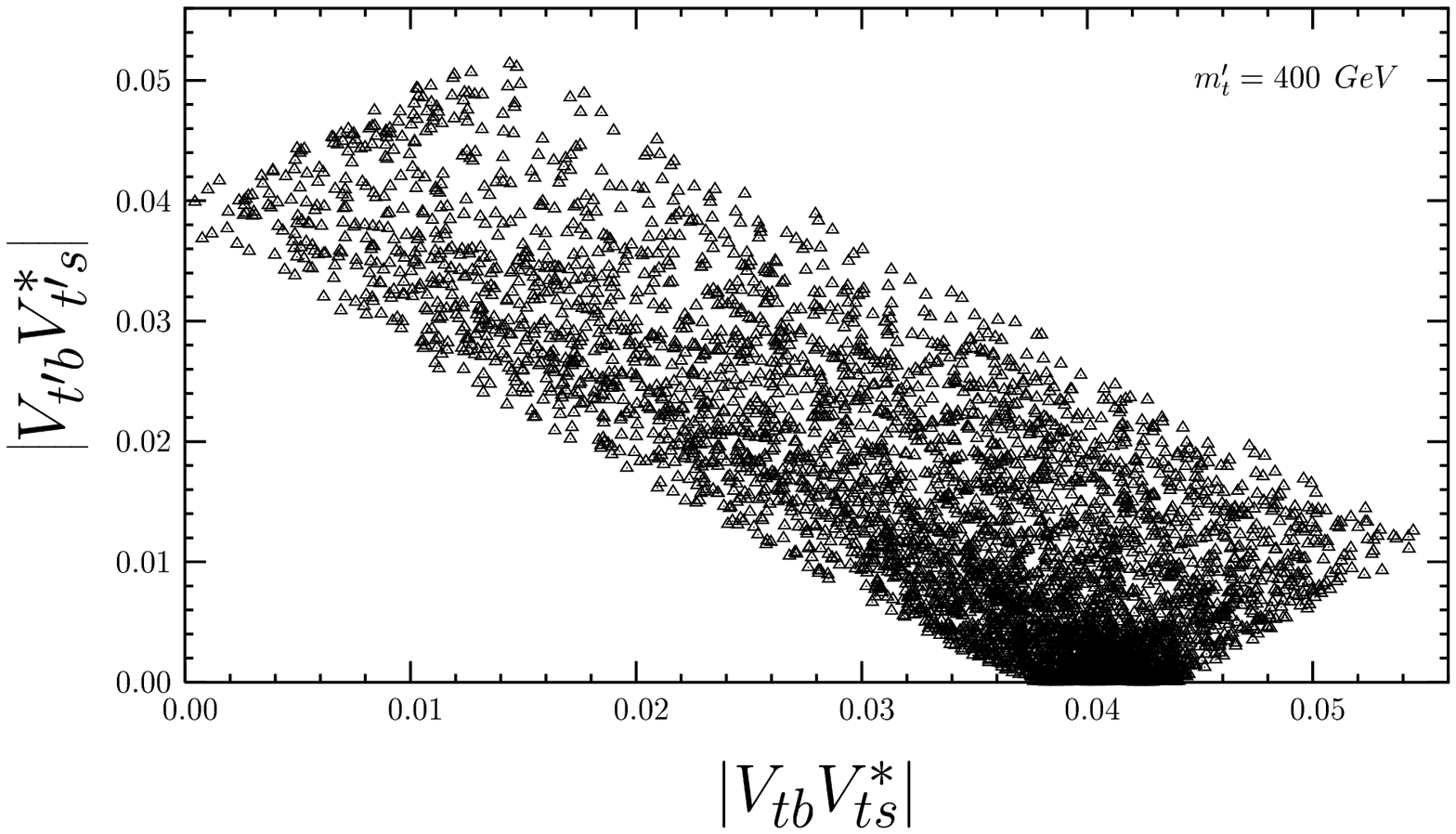}
\vskip 7.8cm
\caption{}
\end{figure}

\begin{figure}
\vskip 2.5 cm
    \includegraphics{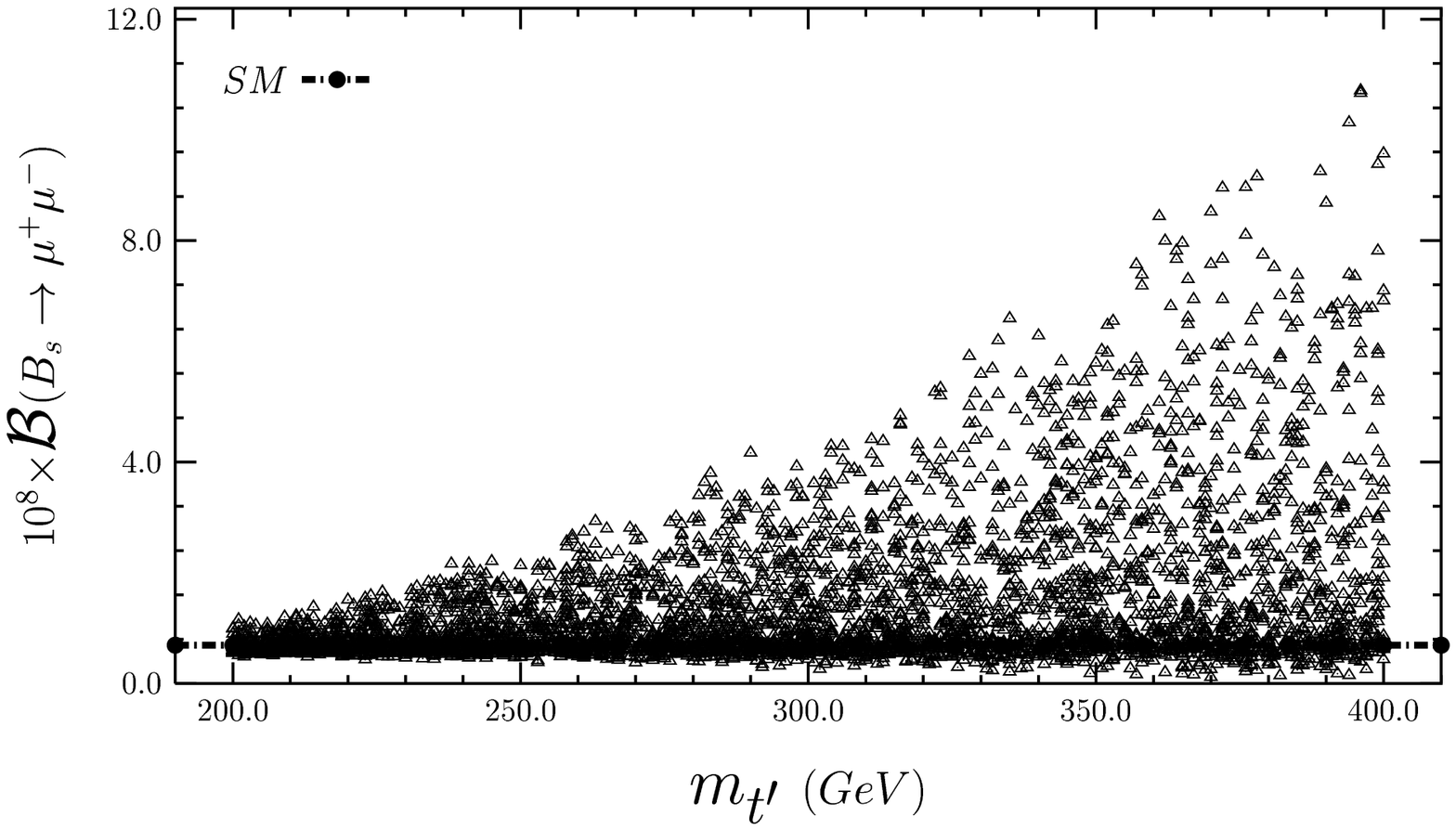}
\vskip 7.8 cm
\caption{}
\end{figure}

\begin{figure}
\vskip 1.5 cm
    \includegraphics{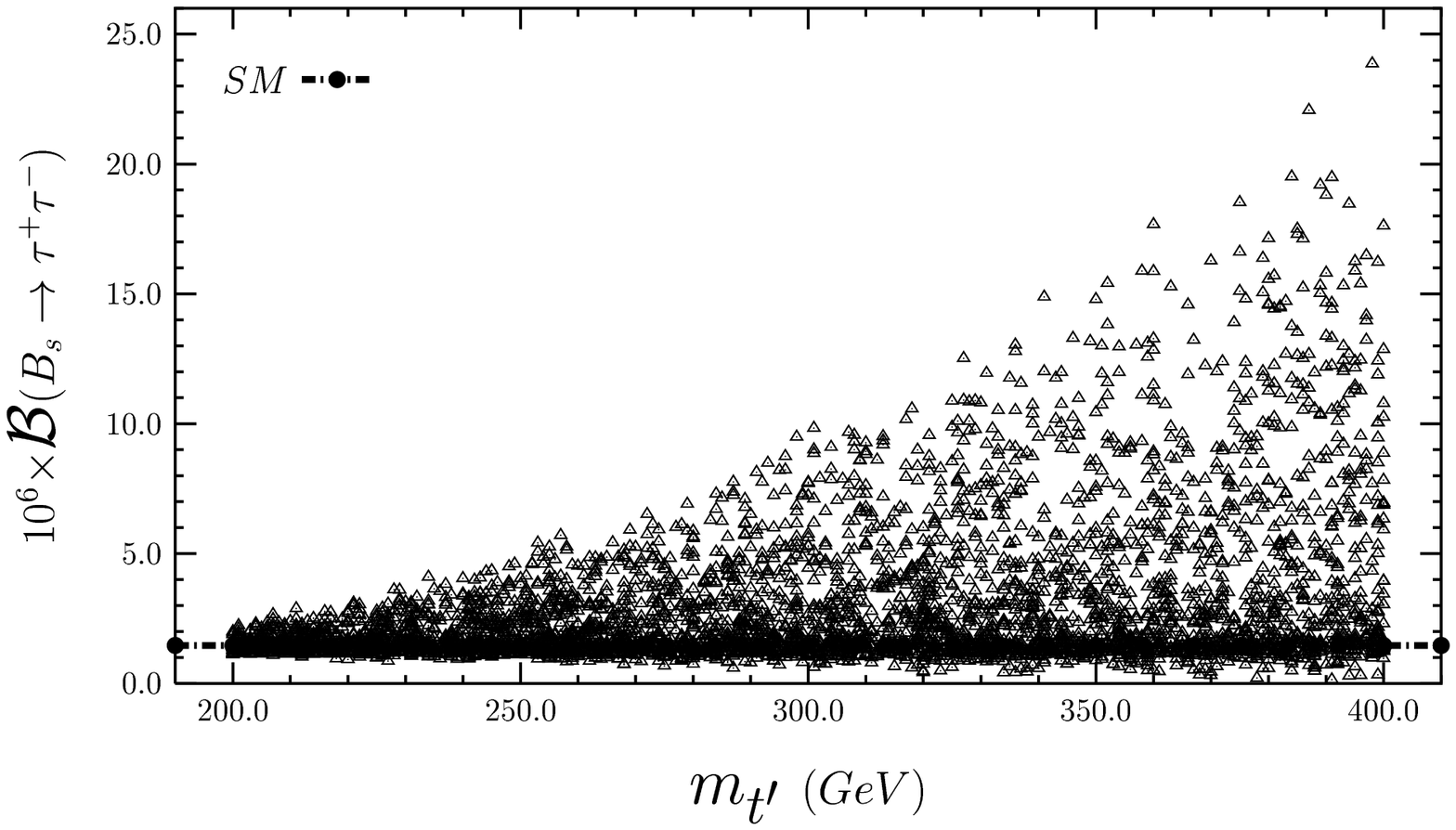}
\vskip 7.8cm
\caption{}
\end{figure}

\begin{figure}
\vskip 2.5 cm
    \includegraphics{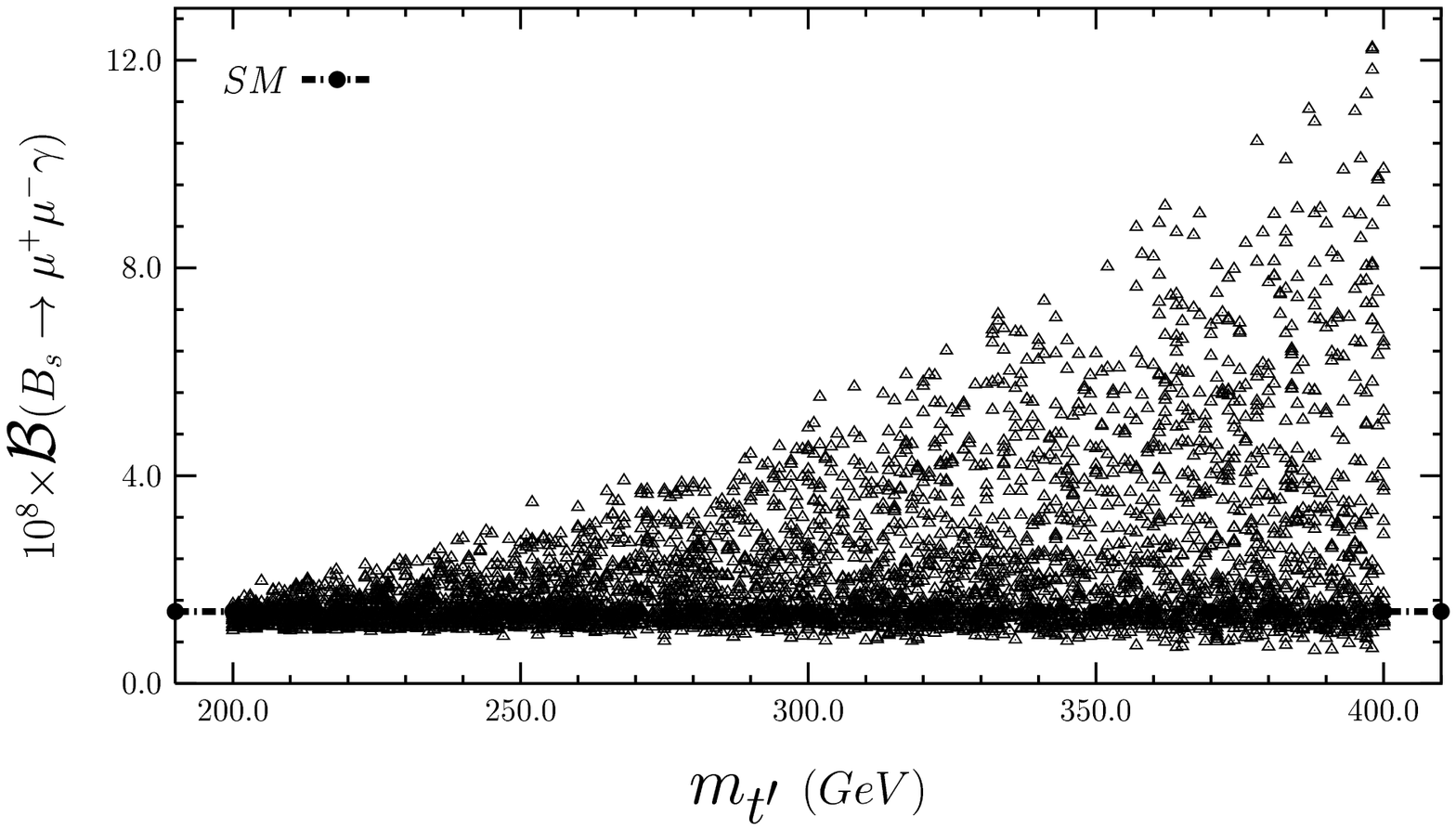}
\vskip 7.8 cm
\caption{}
\end{figure}

\begin{figure}
\vskip 1.5 cm
    \includegraphics{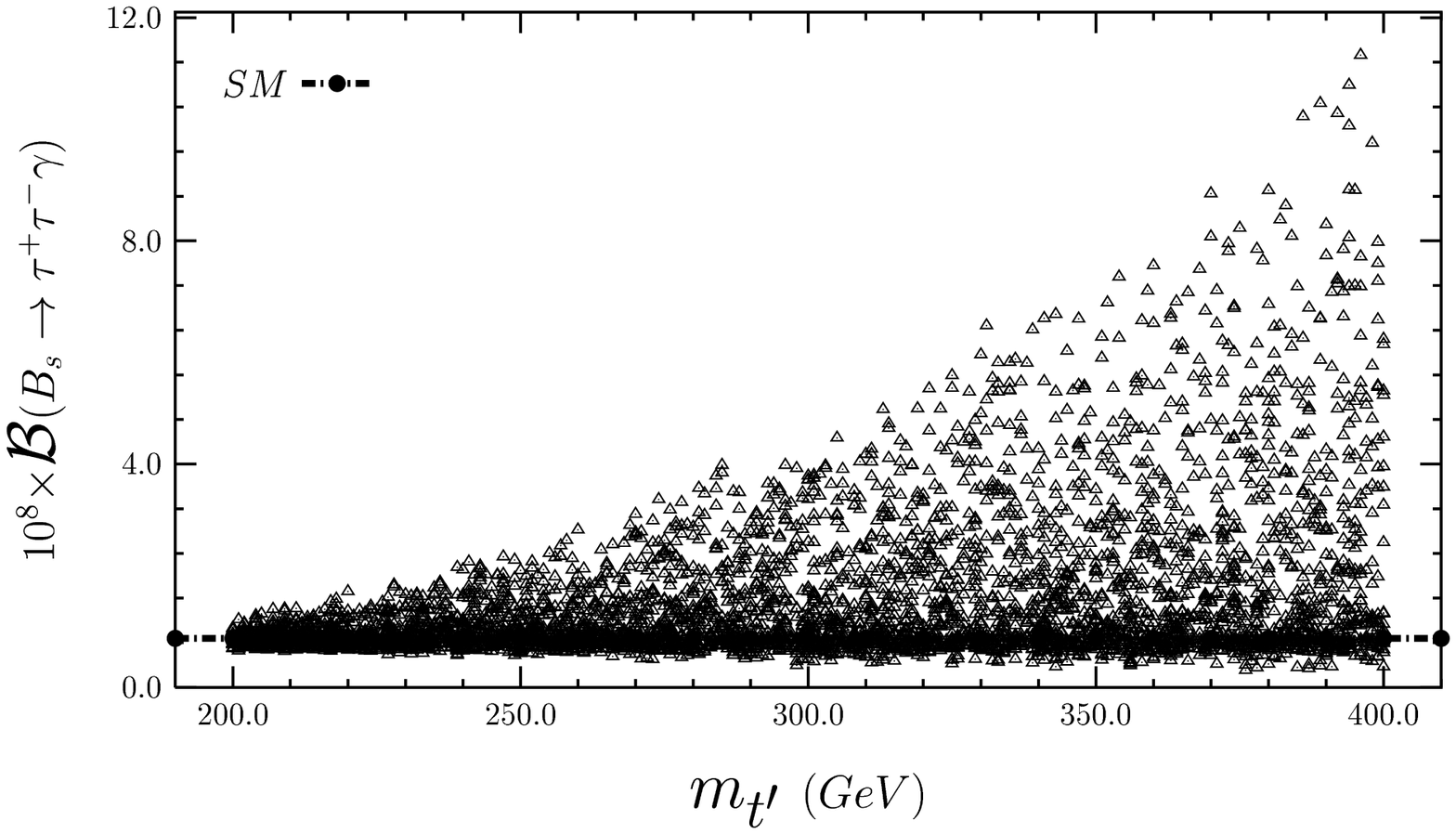}
\vskip 7.8cm
\caption{}
\end{figure}

\end{document}